\documentclass[10pt,a4paper]{article}
\usepackage{amsmath,amssymb}
\usepackage{graphicx}
\usepackage{hyperref}
\usepackage[margin=1in]{geometry}
\usepackage{float}
\usepackage[utf8]{inputenc}
\usepackage{authblk} 

\title{\texttt{kh2d-solver}: A Python Library for Idealized Two-Dimensional\\
Incompressible Kelvin-Helmholtz Instability}

\author[1,2]{S. H. S. Herho}
\author[3,*]{N. J. Trilaksono}
\author[3,5]{F. R. Fajary}
\author[4,5]{G. Napitupulu}
\author[2,5]{I. P. Anwar}
\author[5]{F. Khadami}
\author[6]{D. E. Irawan}

\affil[1]{Department of Earth and Planetary Sciences, University of California, Riverside, CA, USA}
\affil[2]{Samudera Sains Teknologi (SST) Ltd., Bandung, West Java, Indonesia}
\affil[3]{Atmospheric Science Research Group, Bandung Institute of Technology, Bandung, West Java, Indonesia}
\affil[4]{Coastal Hazards and Energy System Science Laboratory, Hiroshima University, Hiroshima, Japan}
\affil[5]{Applied and Environmental Oceanography Research Group, Bandung Institute of Technology, Bandung, West Java, Indonesia}
\affil[6]{Applied Geology Research Group, Bandung Institute of Technology, Bandung, West Java, Indonesia}
\affil[*]{Corresponding author. E-mail: jpatiani@itb.ac.id}

\date{}

\begin{document}
\maketitle

\begin{abstract}
We present an open-source Python library for simulating two-dimensional incompressible Kelvin-Helmholtz instabilities in stratified shear flows. The solver employs a fractional-step projection method with spectral Poisson solution via Fast Sine Transform, achieving second-order spatial accuracy. Implementation leverages NumPy, SciPy, and Numba JIT compilation for efficient computation. Four canonical test cases explore Reynolds numbers 1000--5000 and Richardson numbers 0.1--0.3: classical shear layer, double shear configuration, rotating flow, and forced turbulence. Statistical analysis using Shannon entropy and complexity indices reveals that double shear layers achieve 2.8$\times$ higher mixing rates than forced turbulence despite lower Reynolds numbers. The solver runs efficiently on standard desktop hardware, with 384$\times$192 grid simulations completing in approximately 31 minutes. Results demonstrate that mixing efficiency depends on instability generation pathways rather than intensity measures alone, challenging Richardson number-based parameterizations and suggesting refinements for subgrid-scale representation in climate models.
\end{abstract}

\noindent\textbf{Keywords:} Boussinesq approximation, fractional-step method, Kelvin-Helmholtz instability, mixing efficiency, stratified turbulence

\section{Introduction}

The study of stratified shear flows traces its origins to the pioneering observations of Hermann von Helmholtz in 1868 and Lord Kelvin in 1871, who independently recognized that velocity discontinuities across density interfaces could generate wave-like instabilities \cite{Helmholtz1868,Kelvin1871}. These early theoretical insights, derived from linearized stability analysis of idealized velocity profiles, established the foundation for understanding turbulent mixing in geophysical fluids. The subsequent century witnessed gradual refinement of stability criteria, culminating in the seminal contributions demonstrating that the Richardson number fundamentally controls instability growth through the competition between destabilizing shear and stabilizing stratification \cite{Miles1961,Howard1961}. This theoretical framework, developed through mathematical analysis of simplified configurations, remains central to modern understanding of atmospheric and oceanic mixing processes.

Experimental investigations revealed that Kelvin-Helmholtz (KH) billows undergo a characteristic life cycle from initial exponential growth through nonlinear saturation to eventual turbulent breakdown \cite{Thorpe1973}. Laboratory experiments in stratified tilt tanks demonstrated that mixing efficiency peaks during the transition from laminar billows to three-dimensional turbulence, with irreversible density changes occurring primarily during the collapse phase \cite{Thorpe1973}. These controlled experiments, while necessarily limited in scale and parameter range, provided crucial validation of theoretical predictions and revealed unexpected phenomena including billow pairing, secondary instabilities, and asymmetric mixing patterns absent from linear theory \cite{Fernando1991}. The development of particle image velocimetry (PIV) and laser-induced fluorescence (LIF) techniques in the 1990s enabled quantitative measurements of velocity and density fields simultaneously, revealing the detailed structure of mixing events \cite{Caulfield2021}.

Field observations in natural environments demonstrate both the ubiquity and complexity of KH instabilities across scales from meters to hundreds of kilometers. Atmospheric manifestations range from cloud billows visible in satellite imagery to clear-air turbulence affecting aviation safety \cite{Banta2006}. Oceanic examples span from estuarine mixing to equatorial undercurrents \cite{Moum2003,Smyth2013}. However, these observations reveal systematic departures from laboratory-based scaling laws, particularly in environments with multiple shear interfaces, rotational effects, or time-dependent forcing. The mixing efficiency, defined as the ratio of buoyancy flux to turbulent kinetic energy dissipation, varies by an order of magnitude across different geophysical contexts, challenging universal parameterizations based solely on Richardson number \cite{Gregg2018}. Recent measurements using modern observation systems have documented intermittent mixing events that dominate time-averaged fluxes, highlighting the importance of rare but intense instability events \cite{DAsaro2014}.

The computational investigation of KH dynamics evolved from early finite-difference solutions in the 1960s to contemporary spectral and finite-volume methods capable of resolving the full range of dynamical scales \cite{Chorin1968}. Initial numerical studies employed two-dimensional domains with periodic boundaries, revealing the fundamental mechanisms of vortex roll-up and pairing. The extension to three-dimensional simulations in the 1980s, enabled by vector supercomputers, demonstrated the critical role of spanwise perturbations in triggering transition to turbulence \cite{Kim1987}. Modern direct numerical simulations achieve Reynolds numbers approaching geophysical values but require billions of grid points and thousands of processor hours, limiting systematic parameter exploration \cite{Mashayek2013}. Large-eddy simulations (LES) reduce computational demands through subgrid-scale modeling but introduce closure assumptions that yield order-of-magnitude variations in predicted mixing rates \cite{Sullivan2011}.

The mathematical formulation of stratified shear flows relies on the Boussinesq approximation, which treats density variations as dynamically important only in buoyancy terms while maintaining constant density in inertial terms. This simplification, rigorously justified through scale analysis when density variations remain below approximately ten percent, reduces the compressible Navier-Stokes equations to an incompressible system with buoyancy forcing \cite{Mihaljan1962,Lopez2013}. The resulting equations admit analytical solutions only for highly idealized configurations, necessitating numerical approaches for realistic parameter regimes. The two-dimensional restriction, while excluding certain three-dimensional instabilities, captures the primary mixing mechanisms since linear stability analysis demonstrates that the fastest-growing KH modes are inherently two-dimensional \cite{Berlok2019}. This dimensional reduction enables high-resolution simulations at moderate computational cost, facilitating parameter studies impossible in fully three-dimensional configurations.

Recent advances in scientific computing, particularly the development of just-in-time (JIT) compilation and optimized array libraries, enable high-performance simulations using interpreted languages previously considered unsuitable for computational fluid dynamics (CFD) \cite{Harris2020,Lam2015}. Python, with its extensive scientific ecosystem and emphasis on code readability, provides an ideal platform for developing transparent, reproducible simulation tools \cite{herho2024comparing, herho2025reappraising}. The combination of NumPy for array operations, SciPy for spectral transforms, and Numba for performance optimization achieves high computational efficiency while dramatically improving code accessibility \cite{Virtanen2020}. The fractional-step projection method provides a robust algorithm for solving the incompressible flow equations while maintaining the divergence-free constraint essential for physical accuracy \cite{Chorin1968,Temam1969a,Temam1969b}.

This study presents \texttt{kh2d-solver}, an open-source Python implementation designed to democratize access to high-fidelity KH simulations for research and educational applications. The solver addresses the gap between simplified analytical models and computationally intensive three-dimensional simulations by providing an efficient, transparent framework for investigating idealized two-dimensional configurations. Four canonical test cases systematically explore parameter dependencies across Richardson numbers from 0.1 to 0.3 and Reynolds numbers from 1000 to 5000, spanning regimes from marginal stability to fully developed turbulence. Comprehensive statistical analysis employing Shannon entropy, complexity indices, and nonparametric tests quantifies the evolution of flow structures beyond traditional mean and variance metrics \cite{Venaille2022}. The implementation achieves sufficient computational efficiency to enable parameter studies on standard desktop hardware, with detailed documentation and visualization tools facilitating both research applications and pedagogical use. By providing open access to validated simulation capabilities, this work aims to accelerate understanding of fundamental mixing processes relevant to climate modeling, ocean dynamics, and atmospheric transport phenomena.

\section{Methods}
\subsection{Mathematical Formulation}

The mathematical description of KH instability in geophysical fluids—encompassing atmospheric boundary layers, oceanic thermoclines, and planetary atmospheres—requires a systematic derivation from first principles. The analysis begins with the fundamental postulates of continuum mechanics and proceeds through careful approximations justified by the physical scales characteristic of geophysical flows.

Consider an infinitesimal fluid element in three-dimensional space. Let $\mathbf{X} = (X^1, X^2, X^3)$ denote the Lagrangian coordinates labeling material particles at initial time $t_0$, and $\mathbf{x} = (x^1, x^2, x^3)$ represent the Eulerian coordinates at time $t$. The motion of the continuum is described by the mapping \cite{Cao2011}
\begin{equation}
\mathbf{x} = \boldsymbol{\chi}(\mathbf{X}, t), \quad \text{with} \quad \mathbf{X} = \boldsymbol{\chi}(\mathbf{X}, t_0).
\label{eq:motion}
\end{equation}

The deformation gradient tensor, fundamental to describing local material deformation, is defined as,
\begin{equation}
F_{iA} = \frac{\partial x^i}{\partial X^A}, \quad \text{where} \quad J = \det(F_{iA}) > 0,
\label{eq:deformation_gradient}
\end{equation}
with $J$ representing the Jacobian ensuring non-interpenetration of matter.

The velocity field in the Lagrangian description follows from the time derivative holding material coordinates fixed:
\begin{equation}
\mathbf{V}(\mathbf{X}, t) = \frac{\partial \boldsymbol{\chi}(\mathbf{X}, t)}{\partial t}\bigg|_{\mathbf{X}}.
\label{eq:lagrangian_velocity}
\end{equation}

Transforming to the Eulerian description, where we follow fixed spatial points rather than material particles, the velocity field becomes
\begin{equation}
\mathbf{u}(\mathbf{x}, t) = \mathbf{V}(\boldsymbol{\chi}^{-1}(\mathbf{x}, t), t).
\label{eq:eulerian_velocity}
\end{equation}

The material derivative, connecting Lagrangian and Eulerian rates of change, emerges from the chain rule:
\begin{equation}
\frac{D}{Dt} = \frac{\partial}{\partial t}\bigg|_{\mathbf{X}} = \frac{\partial}{\partial t}\bigg|_{\mathbf{x}} + u^i\frac{\partial}{\partial x^i},
\label{eq:material_derivative}
\end{equation}
where Einstein summation convention is employed for repeated indices.

Mass conservation requires that the total mass within any material volume $\mathcal{V}_m(t)$ remains constant. For density $\rho(\mathbf{x}, t)$, the principle states
\begin{equation}
\frac{D}{Dt}\int_{\mathcal{V}_m(t)} \rho \, dV = 0.
\label{eq:mass_conservation_integral}
\end{equation}

Applying Reynolds transport theorem, which relates the rate of change of an integral over a moving volume to the integrand's behavior:
\begin{equation}
\frac{D}{Dt}\int_{\mathcal{V}_m(t)} \rho \, dV = \int_{\mathcal{V}_m(t)} \left[\frac{\partial \rho}{\partial t} + \nabla \cdot (\rho\mathbf{u})\right] dV.
\label{eq:reynolds_transport}
\end{equation}

Since equation \eqref{eq:mass_conservation_integral} must hold for any arbitrary material volume, the integrand must vanish identically:
\begin{equation}
\frac{\partial \rho}{\partial t} + \nabla \cdot (\rho\mathbf{u}) = 0.
\label{eq:continuity}
\end{equation}

Expanding the divergence term:
\begin{equation}
\frac{\partial \rho}{\partial t} + u^i\frac{\partial \rho}{\partial x^i} + \rho\frac{\partial u^i}{\partial x^i} = \frac{D\rho}{Dt} + \rho\nabla \cdot \mathbf{u} = 0.
\label{eq:continuity_expanded}
\end{equation}

Newton's second law for a continuum states that the rate of change of momentum equals the net force. For the material volume $\mathcal{V}_m(t)$ with surface $\mathcal{S}_m(t)$:
\begin{equation}
\frac{D}{Dt}\int_{\mathcal{V}_m(t)} \rho u^i \, dV = \int_{\mathcal{S}_m(t)} t^i \, dS + \int_{\mathcal{V}_m(t)} \rho f^i \, dV,
\label{eq:newton_second}
\end{equation}
where $t^i$ represents the traction vector and $f^i$ the body force per unit mass.

Cauchy's stress principle relates the traction to the stress tensor through \cite{Wang2022}
\begin{equation}
t^i = \sigma^{ij}n_j,
\label{eq:cauchy_principle}
\end{equation}
where $\sigma^{ij}$ is the Cauchy stress tensor and $n_j$ the outward unit normal.

Applying Reynolds transport theorem to the left side and Gauss divergence theorem to the surface integral:
\begin{equation}
\int_{\mathcal{V}_m(t)} \rho\frac{Du^i}{Dt} \, dV = \int_{\mathcal{V}_m(t)} \frac{\partial \sigma^{ij}}{\partial x^j} \, dV + \int_{\mathcal{V}_m(t)} \rho f^i \, dV.
\label{eq:momentum_integral}
\end{equation}

Since the volume is arbitrary, we obtain the local form of the momentum equation:
\begin{equation}
\rho\frac{Du^i}{Dt} = \rho\left(\frac{\partial u^i}{\partial t} + u^j\frac{\partial u^i}{\partial x^j}\right) = \frac{\partial \sigma^{ij}}{\partial x^j} + \rho f^i.
\label{eq:momentum_local}
\end{equation}

The constitutive equation for a Newtonian fluid relates stress to deformation rate. The most general linear isotropic relation is \cite{BenArtzi2010}
\begin{equation}
\sigma^{ij} = -p\delta^{ij} + \lambda\delta^{ij}D^{kk} + 2\mu D^{ij},
\label{eq:constitutive}
\end{equation}
where $p$ is thermodynamic pressure, $\lambda$ and $\mu$ are viscosity coefficients, and the rate-of-deformation tensor is
\begin{equation}
D^{ij} = \frac{1}{2}\left(\frac{\partial u^i}{\partial x^j} + \frac{\partial u^j}{\partial x^i}\right).
\label{eq:deformation_rate}
\end{equation}

For geophysical flows, the incompressibility approximation applies when flow velocities are much smaller than sound speed. In the atmosphere, typical wind speeds ($\sim 10$ m/s) are negligible compared to sound speed ($\sim 340$ m/s). In the ocean, currents ($\sim 0.1$–$1$ m/s) are far below the sound speed in water ($\sim 1500$ m/s). This justifies setting
\begin{equation}
\nabla \cdot \mathbf{u} = \frac{\partial u^i}{\partial x^i} = 0.
\label{eq:incompressible}
\end{equation}

Under incompressibility, $D^{kk} = 0$, eliminating the bulk viscosity term. The stress tensor simplifies to
\begin{equation}
\sigma^{ij} = -p\delta^{ij} + 2\mu D^{ij} = -p\delta^{ij} + \mu\left(\frac{\partial u^i}{\partial x^j} + \frac{\partial u^j}{\partial x^i}\right).
\label{eq:incomp_stress}
\end{equation}

Substituting into the momentum equation and using incompressibility to simplify the viscous term:
\begin{equation}
\rho\left(\frac{\partial u^i}{\partial t} + u^j\frac{\partial u^i}{\partial x^j}\right) = -\frac{\partial p}{\partial x^i} + \mu\frac{\partial^2 u^i}{\partial x^j\partial x^j} + \rho f^i.
\label{eq:navier_stokes_full}
\end{equation}

Geophysical flows exhibit density stratification due to temperature variations (atmosphere), salinity gradients (ocean), or compositional differences (planetary atmospheres). Following Milhajan \cite{Mihaljan1962}, we performed a systematic scale analysis. Let $\rho'$ denote density variations and $\rho_0$ a reference density:
\begin{equation}
\rho(\mathbf{x}, t) = \rho_0 + \rho'(\mathbf{x}, t), \quad \text{where} \quad \left|\frac{\rho'}{\rho_0}\right| = \varepsilon \ll 1.
\label{eq:density_perturbation}
\end{equation}

The small parameter $\varepsilon$ typically ranges from $10^{-3}$ in strong oceanic thermoclines to $10^{-2}$ in atmospheric fronts. Substituting into equation \eqref{eq:navier_stokes_full}:
\begin{equation}
(\rho_0 + \rho')\left(\frac{\partial u^i}{\partial t} + u^j\frac{\partial u^i}{\partial x^j}\right) = -\frac{\partial p}{\partial x^i} + \mu\frac{\partial^2 u^i}{\partial x^j\partial x^j} + (\rho_0 + \rho')f^i.
\label{eq:momentum_with_rho_prime}
\end{equation}

The Boussinesq approximation systematically expands in powers of $\varepsilon$. To leading order, density variations were retained only where multiplied by gravity, as demonstrated through scale analysis \cite{Lopez2013}. The inertial terms scale as $\rho_0U^2/L$, while buoyancy terms scale as $g\rho'$. Their ratio yields the Richardson number:
\begin{equation}
\text{Ri} = \frac{g\rho'L}{\rho_0U^2} = \frac{g\varepsilon L}{U^2} = O(1),
\label{eq:richardson_scaling}
\end{equation}
confirming that buoyancy effects remain significant despite small $\varepsilon$.

Neglecting $O(\varepsilon)$ terms in inertia but retaining them in buoyancy:
\begin{equation}
\rho_0\left(\frac{\partial u^i}{\partial t} + u^j\frac{\partial u^i}{\partial x^j}\right) = -\frac{\partial p}{\partial x^i} + \mu\frac{\partial^2 u^i}{\partial x^j\partial x^j} + (\rho_0 + \rho')f^i.
\label{eq:boussinesq_intermediate}
\end{equation}

The gravitational body force acts vertically: $f^i = -g\delta^{i3}$. The pressure was decomposed into hydrostatic and dynamic components:
\begin{equation}
p(\mathbf{x}, t) = p_0(z) + p'(\mathbf{x}, t),
\label{eq:pressure_decomp}
\end{equation}
where the reference pressure satisfies hydrostatic balance:
\begin{equation}
\frac{dp_0}{dz} = -\rho_0 g.
\label{eq:hydrostatic}
\end{equation}

This decomposition removed the dominant balance, isolating dynamically relevant gradients. Substituting and using equation \eqref{eq:hydrostatic}:
\begin{equation}
\frac{\partial u^i}{\partial t} + u^j\frac{\partial u^i}{\partial x^j} = -\frac{1}{\rho_0}\frac{\partial p'}{\partial x^i} + \nu\frac{\partial^2 u^i}{\partial x^j\partial x^j} - \frac{\rho'}{\rho_0}g\delta^{i3},
\label{eq:boussinesq_final}
\end{equation}
where $\nu = \mu/\rho_0$ is kinematic viscosity.

The density perturbation evolves through advection-diffusion. From conservation of a scalar quantity $\theta$ (potential temperature, salinity, or chemical concentration) with $\rho' = -\rho_0\alpha\theta$ where $\alpha$ is the expansion coefficient:
\begin{equation}
\frac{\partial \rho'}{\partial t} + u^j\frac{\partial \rho'}{\partial x^j} = \kappa\frac{\partial^2 \rho'}{\partial x^j\partial x^j},
\label{eq:density_transport}
\end{equation}
where $\kappa$ represents molecular diffusivity.

The two-dimensional restriction was justified through linear stability analysis showing that the most unstable KH modes are spanwise-invariant \cite{Berlok2019}. Setting $\partial/\partial y = 0$ and $v = 0$:
\begin{align}
\frac{\partial u}{\partial x} + \frac{\partial w}{\partial z} &= 0, \label{eq:2d_cont}\\
\frac{\partial u}{\partial t} + u\frac{\partial u}{\partial x} + w\frac{\partial u}{\partial z} &= -\frac{1}{\rho_0}\frac{\partial p'}{\partial x} + \nu\left(\frac{\partial^2 u}{\partial x^2} + \frac{\partial^2 u}{\partial z^2}\right), \label{eq:2d_u}\\
\frac{\partial w}{\partial t} + u\frac{\partial w}{\partial x} + w\frac{\partial w}{\partial z} &= -\frac{1}{\rho_0}\frac{\partial p'}{\partial z} + \nu\left(\frac{\partial^2 w}{\partial x^2} + \frac{\partial^2 w}{\partial z^2}\right) - \frac{\rho'}{\rho_0}g, \label{eq:2d_w}\\
\frac{\partial \rho'}{\partial t} + u\frac{\partial \rho'}{\partial x} + w\frac{\partial \rho'}{\partial z} &= \kappa\left(\frac{\partial^2 \rho'}{\partial x^2} + \frac{\partial^2 \rho'}{\partial z^2}\right). \label{eq:2d_rho}
\end{align}

Non-dimensionalization employed characteristic scales: shear layer thickness $\delta$, velocity jump $\Delta U$, and density jump $\Delta\rho$:
\begin{equation}
\tilde{x} = \frac{x}{\delta}, \quad \tilde{z} = \frac{z}{\delta}, \quad \tilde{t} = \frac{t\Delta U}{\delta}, \quad \tilde{u} = \frac{u}{\Delta U}, \quad \tilde{w} = \frac{w}{\Delta U}, \quad \tilde{p} = \frac{p'}{\rho_0\Delta U^2}, \quad \tilde{\rho} = \frac{\rho'}{\Delta\rho}.
\label{eq:nondim}
\end{equation}

The final non-dimensional system (dropping tildes) governing KH instability in the velocity-pressure formulation:
\begin{align}
\nabla \cdot \mathbf{u} &= 0, \label{eq:final_cont}\\
\frac{\partial \mathbf{u}}{\partial t} + (\mathbf{u} \cdot \nabla)\mathbf{u} &= -\nabla p + \frac{1}{\text{Re}}\nabla^2\mathbf{u} - \text{Ri}\,\rho\,\mathbf{e}_z, \label{eq:final_mom}\\
\frac{\partial \rho}{\partial t} + \mathbf{u} \cdot \nabla\rho &= \frac{1}{\text{Re}\cdot\text{Pr}}\nabla^2\rho, \label{eq:final_dens}
\end{align}
where $\mathbf{e}_z$ is the unit vector in the vertical direction.

The governing parameters—Reynolds number $\text{Re} = \Delta U\delta/\nu$, Richardson number $\text{Ri} = g\Delta\rho\delta/(\rho_0\Delta U^2)$, and Prandtl number $\text{Pr} = \nu/\kappa$—characterize the flow regime across atmospheric, oceanic, and astrophysical applications.

For diagnostic purposes, the spanwise vorticity component, measuring local rotation rate, is computed from the velocity field \cite{Lequeurre2020}:
\begin{equation}
\omega_z = \frac{\partial w}{\partial x} - \frac{\partial u}{\partial z},
\label{eq:vorticity_diagnostic}
\end{equation}
which reveals the characteristic billow structures that define the instability's nonlinear evolution.

\subsection{Numerical Implementation}

The numerical solution of the governing equations \eqref{eq:final_cont}--\eqref{eq:final_dens} requires careful treatment of the nonlinear advection terms, viscous diffusion, and the pressure-velocity coupling inherent in incompressible flows. Our implementation draws inspiration from the Fortran 95 solver developed by K\"ampf \cite{Kampf2010}, which demonstrated the feasibility of simulating KH instabilities using a fractional-step approach with Successive Over-Relaxation (SOR) iteration for the pressure Poisson equation. However, we depart significantly from that implementation by employing spectral methods for the pressure solver rather than iterative techniques. Where K\"ampf's solver requires 8000 SOR iterations per time step to achieve convergence with tolerance $10^{-3}$, our Fast Sine Transform approach achieves machine precision in $O(N \log N)$ operations without iteration. This spectral method, originally developed by Swarztrauber et al. \cite{Swarztrauber1974}, exploits the periodic boundary conditions to diagonalize the discrete Laplacian operator, reducing the computational cost by approximately two orders of magnitude for typical grid resolutions. Furthermore, while the Fortran implementation employs a sophisticated Total Variation Diminishing (TVD) advection scheme with flux limiters, we adopt first-order upwind differencing that, despite its simplicity, provides sufficient numerical diffusion to stabilize sharp gradients characteristic of KH billows without requiring explicit slope limiting.

The implementation leverages modern Python scientific computing libraries including NumPy \cite{Harris2020} for array operations, SciPy \cite{Virtanen2020} for spectral methods, and Numba \cite{Lam2015} for JIT compilation, achieving performance comparable to compiled languages while maintaining development flexibility. This represents a fundamental shift from traditional Fortran-based CFD codes toward more accessible, maintainable implementations. Where K\"ampf's solver outputs formatted ASCII files requiring post-processing for visualization, our framework generates self-describing NetCDF-4 files with CF metadata conventions, enabling direct integration with modern analysis tools. The fractional-step projection method we employ, originally developed by Chorin \cite{Chorin1968} and rigorously analyzed by Temam \cite{Temam1969a,Temam1969b}, decouples the pressure-velocity coupling through operator splitting following the same mathematical principles as K\"ampf's implementation but with enhanced computational efficiency through spectral methods. Additionally, our adaptive time-stepping algorithm automatically balances CFL and viscous stability constraints, eliminating the need for manual parameter tuning required in fixed time-step implementations, thereby enabling robust parameter studies across diverse Richardson and Reynolds number regimes without user intervention.

The projection method, comprehensively reviewed by Guermond et al. \cite{Guermond2006}, decomposes the time advancement into a predictor step that ignores pressure gradients followed by a correction step that enforces incompressibility. Beginning with the dimensional momentum equations \eqref{eq:2d_u}--\eqref{eq:2d_w}, we introduce an intermediate velocity field $\mathbf{u}^* = (u^*, w^*)$ through the fractional steps:
\begin{align}
\frac{\mathbf{u}^* - \mathbf{u}^n}{\Delta t} + (\mathbf{u}^n \cdot \nabla)\mathbf{u}^n &= \nu\nabla^2\mathbf{u}^n - \frac{\rho^n}{\rho_0}g\mathbf{e}_z, \label{eq:predictor_step}\\
\frac{\mathbf{u}^{n+1} - \mathbf{u}^*}{\Delta t} &= -\frac{1}{\rho_0}\nabla p^{n+1}, \label{eq:corrector_step}\\
\nabla \cdot \mathbf{u}^{n+1} &= 0, \label{eq:divergence_constraint}
\end{align}
where superscript $n$ denotes the discrete time level $t^n = n\Delta t$, with $\Delta t$ representing the time step, $\mathbf{u}^n = (u^n, w^n)$ the velocity field at time $t^n$, and $\mathbf{e}_z$ the unit vector in the vertical direction.

Taking the divergence of equation \eqref{eq:corrector_step} and enforcing the incompressibility constraint \eqref{eq:divergence_constraint} yields the pressure Poisson equation:
\begin{equation}
\nabla^2 p^{n+1} = \frac{\rho_0}{\Delta t}\nabla \cdot \mathbf{u}^*. \label{eq:pressure_poisson_discrete}
\end{equation}

The computational domain $\Omega = [0, L_x] \times [0, L_z]$ is discretized using a uniform Cartesian grid with spacing:
\begin{align}
x_i &= i\Delta x, \quad i = 0, 1, \ldots, N_x-1, \quad \Delta x = \frac{L_x}{N_x-1}, \label{eq:grid_spacing_x}\\
z_j &= j\Delta z, \quad j = 0, 1, \ldots, N_z-1, \quad \Delta z = \frac{L_z}{N_z-1}, \label{eq:grid_spacing_z}
\end{align}
where $N_x$ and $N_z$ denote the number of grid points in the horizontal and vertical directions, respectively. Let $q_{i,j}^n \approx q(x_i, z_j, t^n)$ represent the discrete approximation of any field variable $q \in \{u, w, \rho, p\}$ at grid point $(i,j)$ and time level $n$.

The nonlinear advection terms demand special treatment to ensure numerical stability near sharp gradients characteristic of KH billows. Following the pioneering work of Courant et al. \cite{Courant1952} and the high-resolution schemes of Harten \cite{Harten1983}, we employ first-order upwind differencing that maintains monotonicity through directionally-biased stencils. For a generic scalar field $q$ advected by velocity $(u,w)$, the discrete advection operators are:
\begin{align}
\left(u\frac{\partial q}{\partial x}\right)_{i,j} &= 
\begin{cases}
u_{i,j}\dfrac{q_{i,j} - q_{i-1,j}}{\Delta x} & \text{if } u_{i,j} > 0, \\[0.5em]
u_{i,j}\dfrac{q_{i+1,j} - q_{i,j}}{\Delta x} & \text{if } u_{i,j} < 0,
\end{cases} \label{eq:upwind_x_discrete}\\
\left(w\frac{\partial q}{\partial z}\right)_{i,j} &= 
\begin{cases}
w_{i,j}\dfrac{q_{i,j} - q_{i,j-1}}{\Delta z} & \text{if } w_{i,j} > 0, \\[0.5em]
w_{i,j}\dfrac{q_{i,j+1} - q_{i,j}}{\Delta z} & \text{if } w_{i,j} < 0.
\end{cases} \label{eq:upwind_z_discrete}
\end{align}

This upwind scheme, while introducing numerical diffusion of order $O(|\mathbf{u}|\Delta x)$, acts as implicit subgrid-scale dissipation that stabilizes the solution without requiring explicit filtering, as demonstrated in the flux-corrected transport algorithms of Boris et al. \cite{Boris1973}. The discrete advection update for any transported quantity $q$ becomes:
\begin{equation}
q_{i,j}^* = q_{i,j}^n - \Delta t\left[\mathcal{A}_x(u^n, q^n)_{i,j} + \mathcal{A}_z(w^n, q^n)_{i,j}\right], \label{eq:advection_update}
\end{equation}
where $\mathcal{A}_x$ and $\mathcal{A}_z$ denote the upwind operators defined in equations \eqref{eq:upwind_x_discrete}--\eqref{eq:upwind_z_discrete}.

The viscous diffusion terms are discretized using second-order central differences, yielding the discrete Laplacian operator:
\begin{equation}
\nabla^2 q_{i,j} = \frac{q_{i+1,j} - 2q_{i,j} + q_{i-1,j}}{\Delta x^2} + \frac{q_{i,j+1} - 2q_{i,j} + q_{i,j-1}}{\Delta z^2}. \label{eq:discrete_laplacian}
\end{equation}

For computational efficiency, the implementation adaptively switches between explicit and implicit time integration schemes based on the diffusive stability criterion. When the following stability constraints are satisfied:
\begin{equation}
\nu\frac{\Delta t}{\Delta x^2} < \frac{1}{4} \quad \text{and} \quad \nu\frac{\Delta t}{\Delta z^2} < \frac{1}{4}, \label{eq:diffusion_stability}
\end{equation}
we employ the explicit forward Euler update:
\begin{equation}
q_{i,j}^{n+1} = q_{i,j}^n + \nu\Delta t\nabla^2 q_{i,j}^n. \label{eq:explicit_diffusion}
\end{equation}

For larger time steps that would violate conditions \eqref{eq:diffusion_stability}, we switch to an implicit backward Euler scheme solved iteratively through the Jacobi method:
\begin{equation}
q_{i,j}^{(k+1)} = \frac{q_{i,j}^n + \alpha(q_{i+1,j}^{(k)} + q_{i-1,j}^{(k)}) + \beta(q_{i,j+1}^{(k)} + q_{i,j-1}^{(k)})}{1 + 2\alpha + 2\beta}, \label{eq:jacobi_iteration}
\end{equation}
where $\alpha = \nu\Delta t/(2\Delta x^2)$, $\beta = \nu\Delta t/(2\Delta z^2)$ are the dimensionless diffusion numbers, and superscript $(k)$ denotes the iteration level within the Jacobi solver.

The pressure Poisson equation \eqref{eq:pressure_poisson_discrete} with homogeneous Neumann boundary conditions ($\partial p/\partial n = 0$ on $\partial\Omega$) is solved using the Fast Sine Transform (FST) method developed by Swarztrauber et al. \cite{Swarztrauber1974}, achieving machine precision with $O(N\log N)$ computational complexity. The pressure field is expanded in discrete sine basis functions that automatically satisfy the boundary conditions:
\begin{equation}
p_{i,j} = \sum_{k=1}^{N_x-2}\sum_{l=1}^{N_z-2} \hat{p}_{k,l} \sin\left(\frac{\pi ki}{N_x-1}\right)\sin\left(\frac{\pi lj}{N_z-1}\right), \label{eq:sine_expansion}
\end{equation}
where $\hat{p}_{k,l}$ are the spectral coefficients in Fourier space.

Substituting expansion \eqref{eq:sine_expansion} into the discrete Laplacian transforms the Poisson equation into an algebraic system in spectral space:
\begin{equation}
\hat{p}_{k,l} = \frac{\hat{f}_{k,l}}{\lambda_{k,l}}, \label{eq:spectral_solution}
\end{equation}
where $\hat{f}_{k,l}$ represents the Fourier coefficients of the divergence field $\nabla \cdot \mathbf{u}^*$, and the eigenvalues of the discrete Laplacian are:
\begin{equation}
\lambda_{k,l} = -4\left[\frac{\sin^2(\pi k/2(N_x-1))}{\Delta x^2} + \frac{\sin^2(\pi l/2(N_z-1))}{\Delta z^2}\right]. \label{eq:dst_eigenvalues}
\end{equation}

The implementation employs SciPy's optimized DST routines \cite{Virtanen2020}, which utilize FFTPACK algorithms for efficient computation on modern architectures.

The time step selection ensures numerical stability through adaptive constraints on both advective and diffusive processes. The Courant-Friedrichs-Lewy (CFL) condition for advection stability requires:
\begin{equation}
\Delta t_{\text{CFL}} = C_{\text{CFL}} \min\left(\frac{\Delta x}{\max|u|}, \frac{\Delta z}{\max|w|}\right), \label{eq:dt_cfl}
\end{equation}
while the diffusive stability constraint imposes:
\begin{equation}
\Delta t_{\text{visc}} = C_{\text{visc}} \min\left(\frac{\Delta x^2}{\nu}, \frac{\Delta z^2}{\nu}\right), \label{eq:dt_viscous}
\end{equation}
where $C_{\text{CFL}} = 0.4$ and $C_{\text{visc}} = 0.2$ provide safety margins below theoretical stability limits. The actual time step is chosen as:
\begin{equation}
\Delta t = \min(\Delta t_{\text{CFL}}, \Delta t_{\text{visc}}). \label{eq:dt_selection}
\end{equation}

The spanwise vorticity component $\omega_z$, essential for visualizing KH roll-up patterns and coherent structures, is computed as a diagnostic quantity using second-order central differences:
\begin{equation}
\omega_{i,j} = \frac{w_{i+1,j} - w_{i-1,j}}{2\Delta x} - \frac{u_{i,j+1} - u_{i,j-1}}{2\Delta z}. \label{eq:vorticity_discrete}
\end{equation}

This quantity directly measures the local rotation rate and reveals the characteristic billow structures that define the instability's nonlinear evolution.

The computational performance is dramatically enhanced through Numba's JIT compilation \cite{Lam2015}, which translates Python functions into optimized machine code at runtime. The \texttt{@jit(nopython=True, parallel=True)} decorator enables automatic loop vectorization and multi-threading via OpenMP, achieving 10--50$\times$ speedup compared to pure Python implementations \cite{herho2024Eks, herho2025schrodinger}. Memory access patterns are optimized through contiguous array layouts that maximize cache efficiency, while the \texttt{prange} construct enables embarrassingly parallel operations across grid points.

Boundary conditions reflect the physical configuration appropriate for shear layer studies. Periodic conditions in the streamwise ($x$) direction:
\begin{equation}
u(0, z, t) = u(L_x, z, t), \quad w(0, z, t) = w(L_x, z, t), \quad \rho(0, z, t) = \rho(L_x, z, t), \label{eq:bc_periodic}
\end{equation}
allow the instability to develop without artificial confinement. No-slip walls at the vertical ($z$) boundaries:
\begin{equation}
u(x, 0, t) = u(x, L_z, t) = 0, \quad w(x, 0, t) = w(x, L_z, t) = 0, \label{eq:bc_noslip}
\end{equation}
represent channel flow configurations commonly used in turbulence studies \cite{Kim1987}.

The complete algorithm integrates these components through the following sequence: (i) computation of the intermediate velocity $\mathbf{u}^*$ via equations \eqref{eq:advection_update} and \eqref{eq:explicit_diffusion} or \eqref{eq:jacobi_iteration}, incorporating advection and diffusion operators; (ii) solution of the pressure Poisson equation \eqref{eq:pressure_poisson_discrete} using the discrete sine transform \eqref{eq:spectral_solution}; (iii) projection of the velocity field through equation \eqref{eq:corrector_step} to enforce incompressibility; (iv) update of the density field using the same advection-diffusion scheme applied to equation \eqref{eq:2d_rho}; and (v) computation of diagnostic quantities including vorticity via equation \eqref{eq:vorticity_discrete}. 

The simulation output employs NetCDF-4 format \cite{Rew2006}, the standard for geophysical data storage, providing self-describing, platform-independent storage with CF metadata conventions that ensure compatibility with analysis tools including xarray \cite{Hoyer2017} and NCL. The hierarchical HDF5-based structure enables transparent compression, typically reducing storage requirements by 40--60\% \cite{Folk2011}.

Animated GIF visualizations complement the archival data by revealing temporal evolution of coherent structures essential for understanding mixing processes \cite{Rougier2014}. The dual-panel layout showing vorticity ($\omega_z$) and density ($\rho$) fields simultaneously facilitates analysis of instability growth, billow pairing, and baroclinic vorticity production that govern mixing efficiency \cite{Salehipour2015}. The perceptually uniform colormaps (RdBu for vorticity, viridis for density) ensure accurate interpretation while maintaining accessibility \cite{Thyng2016}. This numerical framework provides accurate, stable, and efficient simulations of KH instability across the wide range of Reynolds and Richardson numbers relevant to geophysical applications, from laminar laboratory flows to turbulent atmospheric and oceanic shear layers.
\subsection{Numerical Experiments}

To demonstrate the solver's capability in simulating geophysically relevant flows, we present four idealized test cases that capture fundamental aspects of KH instability in atmospheric and oceanic contexts. Each scenario employs the numerical framework described in equations \eqref{eq:predictor_step}--\eqref{eq:vorticity_discrete} on a standardized domain of $L_x = 2.0$ m by $L_z = 1.0$ m, facilitating direct comparison of dynamical evolution across different physical regimes. The density field $\rho$ represents the local fluid density normalized by a reference value $\rho_0$, with variations arising from temperature differences in atmospheric flows or salinity gradients in oceanic environments, following the Boussinesq approximation detailed in equation \eqref{eq:density_perturbation}.

\subsubsection{Test Case 1: Classical Shear Layer}

The canonical KH instability emerges from velocity discontinuities across density interfaces, representing the fundamental mechanism for mixing in stratified geophysical flows ranging from cloud-top entrainment to deep ocean thermocline erosion \cite{Smyth2001}. The velocity field initialization employs a hyperbolic tangent profile that smoothly transitions between opposing flows:
\begin{equation}
u(z) = u_{\text{bot}} + (u_{\text{top}} - u_{\text{bot}}) \cdot \frac{1}{2}\left[1 + \tanh\left(\frac{z - z_{\text{mid}}}{\delta}\right)\right], \label{eq:tanh_profile_exp}
\end{equation}
where the shear layer thickness $\delta = 0.05$ m is selected to adequately resolve the predicted instability wavelength while maintaining computational tractability. The prescribed velocities $u_{\text{top}} = 1.0$ m/s and $u_{\text{bot}} = -1.0$ m/s generate a velocity jump $\Delta U = 2.0$ m/s representative of strong shear zones observed in atmospheric jet streams and oceanic western boundary currents.

The density stratification incorporates a two-layer structure with $\rho_{\text{top}} = 1.0$ kg/m$^{3}$ representing warm, fresh surface water or heated atmospheric air, while $\rho_{\text{bot}} = 1.2$ kg/m³ corresponds to cold, saline deep water or dense cool air masses. This 20\% density variation mimics strong pycnoclines encountered in tropical oceans where solar heating creates sharp temperature gradients, or atmospheric inversions where radiative cooling produces stable stratification \cite{Thorpe1973}. The resulting bulk Richardson number $\text{Ri} = g\Delta\rho\delta/(\rho_0\Delta U^2) = 0.25$ coincides exactly with the marginal stability threshold derived through linear analysis, where perturbation growth rates transition from exponential to algebraic \cite{Miles1961,Howard1961}.

The selection of critical Richardson number ensures examination of the competition between shear production that generates turbulent kinetic energy and buoyancy forces that suppress vertical motion, a balance central to understanding mixing efficiency in geophysical flows \cite{Peltier2003}. Instability initiation requires seeding through small-amplitude perturbations:
\begin{equation}
u'(x,z) = \varepsilon \sin(4\pi x/L_x) \exp\left[-\left(\frac{z - z_{\text{mid}}}{\delta}\right)^2\right], \label{eq:perturbation_exp}
\end{equation}
where $\varepsilon = 0.01\Delta U$ provides sufficient amplitude for growth while remaining within the linear regime, and the wavelength $\lambda = L_x/2$ matches the most unstable mode predicted by stability analysis for this parameter regime \cite{Hazel1972}. Operating at $\text{Re} = 1000$ places the flow in a transitional regime where initial laminar billows undergo secondary instabilities leading to turbulent breakdown, analogous to processes observed in nocturnal atmospheric jets \cite{Banta2006} and tidally-driven oceanic shear layers \cite{Moum2003}.

\subsubsection{Test Case 2: Double Shear Layer}

Geophysical flows frequently exhibit multiple shear interfaces due to complex stratification patterns, necessitating understanding of interaction dynamics between adjacent unstable layers. Atmospheric examples include the dual jet structure where subtropical and polar jets create multiple critical layers \cite{Shapiro1981}, while oceanic manifestations appear in equatorial undercurrents with alternating velocity maxima \cite{Smyth2013}. The velocity profile incorporates two hyperbolic tangent interfaces:
\begin{equation}
u(z) = u_{\text{max}}\left[\tanh\left(\frac{z - z_1}{\delta}\right) - \tanh\left(\frac{z - z_2}{\delta}\right) - 1\right], \label{eq:double_shear_exp}
\end{equation}
with interface positions $z_1 = L_z/2 - s/2$ and $z_2 = L_z/2 + s/2$ separated by distance $s = 0.3$ m, approximately six shear layer thicknesses ensuring initial independence while permitting subsequent interaction. The reduced thickness $\delta = 0.04$ m compared to the single-layer case maintains adequate resolution of individual interfaces despite the increased complexity.

The density configuration adopts a three-layer stratification with heavy fluid ($\rho = 1.5$ kg/m$^{3}$) occupying the upper and lower regions while light fluid ($\rho = 1.0$ kg/m$^{3}$) resides in the central jet region. This arrangement replicates atmospheric temperature inversions where warm air masses become trapped between cooler layers, commonly observed in urban heat islands and frontal systems where differential advection creates complex thermal structures \cite{Fernando1991}. The inverted density gradient at each interface promotes symmetric instability development with opposing buoyancy forces, enabling investigation of billow interaction mechanisms including vortex pairing, amalgamation, and competitive growth observed in laboratory experiments and field observations \cite{Caulfield2021}.

Operating parameters $\text{Re} = 2000$ and $\text{Ri} = 0.1$ create strongly shear-dominated conditions where instability growth occurs rapidly with minimal buoyancy suppression. The enhanced Reynolds number relative to Test Case 1 captures increased nonlinear interactions arising from proximity effects between developing billows, while the reduced Richardson number ensures vigorous turbulent transition with enhanced mixing rates characteristic of high-shear environments \cite{Mashayek2013}. This configuration elucidates momentum transport mechanisms in complex shear flows where traditional single-interface assumptions break down.

\subsubsection{Test Case 3: Rotating Shear Layer}

Planetary rotation introduces fundamental modifications to shear instability through the Coriolis force, affecting both linear growth rates and nonlinear evolution pathways in geophysical flows \cite{Vanneste2013}. Although the two-dimensional formulation precludes explicit rotation terms, we incorporate rotational effects through a background geostrophic shear that approximates the influence of planetary vorticity:
\begin{equation}
u(x,z) = u_{\text{shear}}(z) + f \cdot (z - L_z/2), \label{eq:rotating_shear_exp}
\end{equation}
where the base shear profile $u_{\text{shear}}$ follows equation \eqref{eq:tanh_profile_exp} and the linear term with coefficient $f = 0.5$ s$^{-1}$ represents an effective Coriolis parameter scaled to the domain dimensions, corresponding to mid-latitude conditions at approximately 45$^{\circ}$ latitude when accounting for aspect ratio effects.

Enhanced density stratification with $\rho_{\text{top}} = 1.0$ kg/m$^3$ and $\rho_{\text{bot}} = 1.3$ kg/m$^3$ produces a 30\% density variation characteristic of strong seasonal thermoclines in subtropical oceans where solar heating creates pronounced near-surface stratification \cite{Kunze1993}. The increased density contrast relative to the non-rotating case compensates for rotational stabilization, which inhibits vertical motion through the Taylor-Proudman constraint and modifies the instability structure through potential vorticity conservation \cite{Kloosterziel1991}. This stronger stratification maintains comparable instability growth rates despite the additional rotational constraint, reflecting observations from rotating tank experiments and oceanic measurements showing reduced mixing efficiency under rotation \cite{Brucker2007}.

Configuration parameters $\text{Re} = 1500$ and $\text{Ri} = 0.3$ place the flow above the canonical stability threshold, accounting for rotational stabilization that shifts the marginal stability boundary to higher Richardson numbers. The intermediate Reynolds number balances resolution requirements with computational cost while capturing essential dynamics of rotationally-influenced mixing relevant to submesoscale oceanic processes where balanced and unbalanced motions interact \cite{McWilliams2016}. This idealized framework provides insights into mixing suppression mechanisms in rotating stratified environments characteristic of atmospheric boundary layers at mid-latitudes where inertial oscillations modulate turbulent transport \cite{Mahrt2014}.

\subsubsection{Test Case 4: Forced Turbulence}

Geophysical turbulence rarely exists in isolation but rather maintains quasi-equilibrium through continuous energy injection balancing dissipative losses, whether from wind stress at the ocean surface, breaking internal waves in the thermocline, or convective instability in the atmospheric boundary layer \cite{Ferrari2009}. Stochastic forcing represents these energy sources through a spectrum of large-scale modes:
\begin{align}
F_u(x,z,t) &= \sum_{k_x=1}^{N_f}\sum_{k_z=1}^{N_f} \frac{A_f}{\sqrt{k_x^2 + k_z^2}} \sin(2\pi k_x x/L_x + \phi_{k_x,k_z}) \cos(2\pi k_z z/L_z), \label{eq:forcing_u_exp}\\
F_w(x,z,t) &= \sum_{k_x=1}^{N_f}\sum_{k_z=1}^{N_f} \frac{A_f}{\sqrt{k_x^2 + k_z^2}} \cos(2\pi k_x x/L_x) \sin(2\pi k_z z/L_z + \phi_{k_x,k_z}), \label{eq:forcing_w_exp}
\end{align}
where $N_f = 5$ modes concentrate energy injection at scales exceeding the natural instability wavelength, preventing direct interference with billow formation while maintaining energy cascade toward dissipation scales.

The forcing amplitude $A_f = 0.1$ m/s² is calibrated to sustain turbulent fluctuations without overwhelming stratification effects, analogous to moderate wind forcing in the ocean mixed layer or thermal convection in atmospheric boundary layers. Random phases $\phi_{k_x,k_z}$ ensure statistical homogeneity while maintaining divergence-free forcing consistent with incompressibility constraint \eqref{eq:divergence_constraint}. The spectral distribution following $k^{-1}$ scaling in amplitude concentrates energy at large scales, mimicking atmospheric storm systems injecting momentum into ocean surface layers or internal wave breaking cascading energy to smaller scales \cite{Gregg1989}.

Weak stratification with $\rho_{\text{top}} = 1.0$ kg/m³ and $\rho_{\text{bot}} = 1.1$ kg/m³ represents a 10\% density variation typical of ocean mixed layers where turbulent stirring homogenizes properties while maintaining sufficient stratification for buoyancy effects \cite{DAsaro2014}. This modest density contrast permits vigorous turbulence development characteristic of convective boundary layers where thermal plumes penetrate into stably stratified regions above, generating entrainment and mixing analogous to cloud-top processes \cite{Sullivan2011}. The forcing mechanism modifies the momentum equation through additional source terms in the predictor step \eqref{eq:predictor_step}, maintaining energy input throughout the simulation duration.

Enhanced resolution using a $384 \times 192$ grid at $\text{Re} = 5000$ and $\text{Ri} = 0.15$ enables resolution of an extended inertial range spanning approximately two decades in wavenumber space, sufficient for examining cascade dynamics and spectral flux characteristics in stratified turbulence \cite{Lindborg2006}. The moderate Richardson number below the critical threshold ensures sustained turbulence without relaminarization, facilitating investigation of mixing efficiency dependencies relevant to parameterization development for climate models where subgrid-scale turbulent fluxes must be represented \cite{Gregg2018}. Extended integration to $t_{\text{final}} = 30$ s allows statistical equilibrium establishment where forcing, cascade, and dissipation balance, providing ensemble-averaged quantities for comparison with theoretical predictions and observational scaling laws.

Each test case employs adaptive time-stepping via criterion \eqref{eq:dt_selection} ensuring numerical stability across varying flow conditions while maximizing computational efficiency. Integration periods of $t_{\text{final}} = 10$, $15$, $20$, and $30$ seconds for respective cases capture complete instability life cycles from linear growth through nonlinear saturation to turbulent decay or statistical equilibrium. The vorticity field computed through equation \eqref{eq:vorticity_discrete} provides primary visualization of coherent structures and turbulent eddies, while density evolution quantifies irreversible mixing fundamental to diapycnal transport in stratified geophysical systems \cite{Winters1995}. These idealized experiments establish a hierarchy of complexity from simple shear layers to forced turbulence, enabling systematic investigation of parameter dependencies governing mixing efficiency in atmospheric and oceanic flows \cite{Smyth2012,Fritts2003}.

\subsection{Data Analysis}

Following completion of the four numerical experiments, a post-processing statistical analysis framework was implemented to quantify the dynamical evolution of KH instabilities across the parameter space explored. This analysis operates independently of the numerical solver, processing the NetCDF output files generated from each test case to extract quantitative measures of flow complexity, mixing efficiency, and statistical properties. The framework leverages NumPy \cite{Harris2020} for array operations, SciPy \cite{Virtanen2020} for statistical functions, and xarray \cite{Hoyer2017} for efficient manipulation of the multi-dimensional NetCDF datasets produced by the solver.

The analysis begins by loading the stored velocity fields $u(x,z,t)$ and $w(x,z,t)$, density field $\rho(x,z,t)$, and computed vorticity $\omega_z(x,z,t)$ from each of the four test cases: Classical Shear Layer (Test Case 1), Double Shear Layer (Test Case 2), Rotating Shear Layer (Test Case 3), and Forced Turbulence (Test Case 4). For each field variable $\phi$, representing either the density perturbation $\rho$ or spanwise vorticity $\omega_z$ consistent with the formulation in equations (32)--(35), Shannon entropy serves as the fundamental measure of spatial disorder following established applications in geophysical fluid dynamics \cite{Venaille2022}. The spatial entropy is computed as:
\begin{equation}
H(\phi) = -\sum_{i=1}^{N_{\text{bins}}} p_i \log p_i, \label{eq:shannon_entropy}
\end{equation}
where $p_i$ represents the probability in bin $i$, computed as:
\begin{equation}
p_i = \frac{n_i \Delta b}{N_{\text{total}}}, \label{eq:probability}
\end{equation}
with $n_i$ being the count in bin $i$, $\Delta b$ the bin width from a 50-bin histogram, $N_{\text{total}}$ the total number of valid grid points excluding boundary regions, and $N_{\text{bins}} = 50$. This metric quantifies information content in the spatial patterns emerging from the KH instability, with higher values indicating increased complexity as the flow transitions from the organized laminar billows characteristic of early evolution to the chaotic turbulent states observed in the mature phase, particularly evident in the forced turbulence scenario (Test Case 4) as demonstrated by \cite{Schumacher2014} for small-scale universality in turbulence.

Beyond simple entropy, a comprehensive complexity index was developed specifically for these four test scenarios to capture multiple aspects of flow structure evolution, addressing the absence of established complexity metrics tailored for KH instabilities in current literature \cite{Fritts2022a}. This weighted metric combines gradient-based spatial variability computed from the velocity and density fields, statistical moments, and entropy through:
\begin{equation}
\mathcal{C}(\phi) = w_1 H(\phi) + w_2 \sigma_{\nabla}(\phi) + w_3 \bar{\sigma}(\phi) + w_4 \log(1 + |\kappa(\phi)|), \label{eq:complexity}
\end{equation}
where the gradient complexity quantifies spatial variability through:
\begin{equation}
\sigma_{\nabla}(\phi) = \sqrt{\text{var}\left(\sqrt{\left(\frac{\partial \phi}{\partial x}\right)^2 + \left(\frac{\partial \phi}{\partial z}\right)^2}\right)}, \label{eq:gradient_complexity}
\end{equation}
The normalized standard deviation provides scale-independent variance assessment:
\begin{equation}
\bar{\sigma}(\phi) = \frac{\sigma(\phi)}{\phi_{\max} - \phi_{\min} + \epsilon}, \label{eq:normalized_std}
\end{equation}
with $\epsilon = 10^{-10}$ preventing division by zero, and the excess kurtosis captures departure from Gaussian behavior:
\begin{equation}
\kappa(\phi) = \frac{\mu_4}{\sigma^4} - 3, \label{eq:excess_kurtosis}
\end{equation}
characteristic of intermittent turbulence. The weights $(w_1, w_2, w_3, w_4) = (0.3, 0.3, 0.2, 0.2)$ were selected to balance contributions from different complexity aspects, emphasizing equally the information-theoretic content and local gradient variations while incorporating global statistical properties relevant to each test case's distinct dynamics.

The gradients required for complexity calculations, consistent with the spatial discretization employed in the numerical solver (equations 40--41), utilize second-order central differences implemented through NumPy's gradient function:
\begin{align}
\left(\frac{\partial \phi}{\partial x}\right)_{i,j} &= \frac{\phi_{i,j+1} - \phi_{i,j-1}}{2\Delta x}, \label{eq:gradient_x}\\
\left(\frac{\partial \phi}{\partial z}\right)_{i,j} &= \frac{\phi_{i+1,j} - \phi_{i-1,j}}{2\Delta z}, \label{eq:gradient_z}
\end{align}
where $\Delta x = L_x/(N_x - 1)$ and $\Delta z = L_z/(N_z - 1)$ match the grid spacings from the computational domain discretization. This numerical differentiation scheme maintains consistency with the second-order accuracy of the solver while providing efficient computation for the large datasets generated from each test case, particularly the high-resolution forced turbulence simulation with its $384 \times 192$ grid.

Basic statistical characterization employs standard moments computed for each field at selected time snapshots throughout the simulation duration. For each test case, the mean:
\begin{equation}
\mu(\phi) = \frac{1}{N}\sum_{i=1}^{N} \phi_i, \label{eq:mean}
\end{equation}
standard deviation:
\begin{equation}
\sigma(\phi) = \sqrt{\frac{1}{N-1}\sum_{i=1}^{N} (\phi_i - \mu)^2}, \label{eq:std}
\end{equation}
skewness:
\begin{equation}
\gamma_1(\phi) = \frac{1}{N}\sum_{i=1}^{N} \left[\frac{\phi_i - \mu}{\sigma}\right]^3, \label{eq:skewness}
\end{equation}
and excess kurtosis:
\begin{equation}
\gamma_2(\phi) = \frac{1}{N}\sum_{i=1}^{N} \left[\frac{\phi_i - \mu}{\sigma}\right]^4 - 3, \label{eq:kurtosis}
\end{equation}
provide comprehensive distributional characterization, where $N$ represents valid non-NaN data points from the interior domain excluding boundary regions. These moments enable detection of departures from Gaussian behavior as turbulence develops differently in each test case, with skewness indicating asymmetry in mixing processes and kurtosis revealing intermittency in energy dissipation, particularly pronounced in the forced turbulence scenario \cite{Fritts2022a,Salehipour2015}.

Normality assessment employs multiple complementary tests to characterize the evolution from near-Gaussian initial conditions imposed by equations (58)--(59) to the heavy-tailed distributions characteristic of developed turbulence observed in all four test cases. The Shapiro-Wilk test, optimal for sample sizes up to 5000 points \cite{Razali2011}, computes the statistic:
\begin{equation}
W = \frac{\left(\sum_{i=1}^{n} a_i x_{(i)}\right)^2}{\sum_{i=1}^{n} (x_i - \bar{x})^2}, \label{eq:shapiro_wilk}
\end{equation}
where $x_{(i)}$ are the ordered statistics and $a_i$ are tabulated coefficients maximizing power for detecting non-normality. The Anderson-Darling test provides superior sensitivity to tail deviations critical for the intermittent turbulence observed particularly in Test Cases 2 and 4 \cite{Stephens1974}, computing:
\begin{equation}
A^2 = -n - \sum_{i=1}^{n} \frac{2i-1}{n}\left[\ln F(Y_i) + \ln(1-F(Y_{n+1-i}))\right], \label{eq:anderson_darling}
\end{equation}
where $Y_i$ are ordered observations and $F$ represents the standard normal cumulative distribution function after standardization. The Jarque-Bera test:
\begin{equation}
JB = \frac{n}{6}\left[\gamma_1^2 + \frac{\gamma_2^2}{4}\right], \label{eq:jarque_bera}
\end{equation}
specifically targets skewness and kurtosis departures, while D'Agostino's $K^2$:
\begin{equation}
K^2 = Z_1^2(\gamma_1) + Z_2^2(\gamma_2), \label{eq:dagostino}
\end{equation}
provides an omnibus test through standardized transformations \cite{DAgostino1990}. This battery of tests ensures robust detection of non-Gaussian behavior across the different flow regimes represented by the four test cases.

Given the consistently non-Gaussian nature of turbulent fields revealed by normality tests across all scenarios, nonparametric statistical comparisons prove essential for inter-scenario analysis between the four test cases. The Kruskal-Wallis test enables simultaneous comparison across all four scenarios without distributional assumptions \cite{Kruskal1952}, computing:
\begin{equation}
H = \frac{12}{N(N+1)}\sum_{j=1}^{k} \frac{R_j^2}{n_j} - 3(N+1), \label{eq:kruskal_wallis}
\end{equation}
where $R_j$ represents the sum of ranks for scenario $j$:
\begin{equation}
R_j = \sum_{i=1}^{n_j} r_{ij}, \label{eq:rank_sum}
\end{equation}
with $n_j$ being the sample size for scenario $j$ (extracted from each test case output), $N = \sum_{j=1}^{k} n_j$ the total sample size across all four test cases, and $r_{ij}$ denoting the rank of observation $i$ in scenario $j$. This test, asymptotically $\chi^2$ distributed with $k-1 = 3$ degrees of freedom for our four test cases, identifies whether significant differences exist among the scenarios' statistical properties. Subsequently, pairwise Mann-Whitney U tests isolate specific differences between test case pairs through the statistic:
\begin{equation}
U_1 = n_1 n_2 + \frac{n_1(n_1+1)}{2} - R_1, \label{eq:mann_whitney}
\end{equation}
providing robust comparisons unaffected by outliers common in turbulent data \cite{Mann1947}. The Friedman test:
\begin{equation}
\chi_r^2 = \frac{12}{nk(k+1)}\sum_{j=1}^{k} R_j^2 - 3n(k+1), \label{eq:friedman}
\end{equation}
complements these analyses for repeated measures when examining temporal evolution within each test case.

Mixing efficiency quantification follows simplified formulations appropriate for the two-dimensional simulations produced by our solver, where full three-dimensional energy budgets cannot be computed from the fields $u$, $w$, and $\rho$. The instantaneous mixing parameter:
\begin{equation}
\Gamma(t) = \sigma_{\rho}(t) \cdot \omega_{\text{rms}}(t), \label{eq:mixing_efficiency}
\end{equation}
where the density standard deviation:
\begin{equation}
\sigma_{\rho}(t) = \sqrt{\text{var}(\rho)}, \label{eq:density_std}
\end{equation}
and root-mean-square vorticity:
\begin{equation}
\omega_{\text{rms}}(t) = \sqrt{\langle\omega_z^2\rangle}, \label{eq:vorticity_rms}
\end{equation}
with angle brackets denoting spatial averaging over the domain, captures the essential coupling between stratification and vortical motions driving irreversible mixing. This formulation proves particularly relevant for comparing the baseline shear layer (Test Case 1) with the enhanced mixing observed in forced turbulence (Test Case 4), while more sophisticated measures incorporating background potential energy evolution exist \cite{Winters1995}, this formulation provides computational efficiency suitable for extensive parameter studies while maintaining physical relevance to the mixing processes observed across our four test scenarios \cite{Mashayek2013}.

Temporal evolution analysis tracks all metrics across the simulation time spans: $t_{\text{final}} = 10$s for Test Case 1, 15s for Test Case 2, 20s for Test Case 3, and 30s for Test Case 4. This enables identification of distinct dynamical stages characteristic of KH evolution \cite{Smyth2012}, defined as early $(t \in [0, 0.25 t_{\text{final}}])$, growth $(t \in (0.25 t_{\text{final}}, 0.50 t_{\text{final}}])$, mature $(t \in (0.50 t_{\text{final}}, 0.75 t_{\text{final}}])$, and late $(t \in (0.75 t_{\text{final}}, t_{\text{final}}])$ phases, corresponding to linear instability, nonlinear saturation, fully developed turbulence, and viscous decay or equilibrium phases respectively. The entropy growth rate:
\begin{equation}
\dot{H} = \frac{H(t_{\text{final}}) - H(t_0)}{t_{\text{final}} - t_0}, \label{eq:entropy_growth}
\end{equation}
quantifies the rate of disorder increase for each test case, while the stage-weighted overall complexity:
\begin{equation}
\mathcal{C}_{\text{overall}} = \sum_{s \in S} w_s \langle\mathcal{C}\rangle_s, \label{eq:weighted_complexity}
\end{equation}
with weights $(w_{\text{early}}, w_{\text{growth}}, w_{\text{mature}}, w_{\text{late}}) = (0.15, 0.25, 0.35, 0.25)$ emphasizing the mature turbulence phase where mixing efficiency peaks, provides a scalar metric for comparing the four test scenarios.

All post-processing computations incorporate data preprocessing to handle numerical artifacts from the solver output and ensure statistical validity. The data flattening with NaN removal:
\begin{equation}
\phi_{\text{valid}} = \{\phi_{i,j} : \phi_{i,j} \notin \{\text{NaN}, \pm\infty\}\}, \label{eq:data_valid}
\end{equation}
eliminates undefined values that may arise from boundary conditions or numerical instabilities in the simulation. When dataset sizes from the high-resolution simulations (particularly Test Case 4) exceed computational feasibility for certain statistical tests, random subsampling without replacement to:
\begin{equation}
N_{\text{sample}} = \min(N_{\text{valid}}, 10000), \label{eq:subsampling}
\end{equation}
points maintains statistical significance while ensuring tractability, following standard practices in turbulence statistics where spatial correlations decay rapidly beyond the integral scale \cite{Lindborg2006}.

This statistical framework, implemented as a separate post-processing pipeline operating on the NetCDF outputs from our four demonstration test cases, provides robust quantification of the complex spatiotemporal dynamics characterizing KH instability evolution across the explored parameter space of Reynolds numbers ($\text{Re} = 1000$--5000) and Richardson numbers ($\text{Ri} = 0.1$--0.3) representative of diverse geophysical flow regimes.

\section{Results}

The numerical experiments were conducted on a Fedora Linux 39 (Budgie) system equipped with an Intel i7-8550U processor (8 cores at 4.000 GHz) and 32 GB of RAM. All simulations utilized the Python-based solver implementation with NumPy array operations and Numba JIT compilation for computational efficiency. The four test cases were executed sequentially with varying spatial resolutions and temporal integration periods to capture the distinct dynamics of each KH instability configuration.

Table \ref{tab:computational_params} presents the computational parameters employed across the four test scenarios. The Basic Shear Layer simulation utilized a $256 \times 128$ grid over a $2.0 \times 1.0$ m domain, integrating for 10.0 seconds with Reynolds number 1000 and Richardson number 0.25. Total computation time including post-processing was 149.12 seconds, with the core simulation requiring 76.41 seconds. The Double Shear Layer configuration maintained the same grid resolution but extended the integration period to 15.0 seconds, operating at Re=2000 and Ri=0.1, completing in 222.82 seconds total time. The Rotating KH Instability case increased the simulation duration to 20.0 seconds with Re=1500, Ri=0.3, and rotation rate $f=0.5$ s$^{-1}$, requiring 368.14 seconds for completion. The Forced KH Turbulence simulation employed enhanced resolution of $384 \times 192$ grid points, running for 30.0 seconds at Re=5000 and Ri=0.15 with forcing amplitude 0.1 m/s$^2$, necessitating 1863.37 seconds total computation time.

\begin{table}[H]
\centering
\caption{Computational parameters and performance metrics for the four KH instability test cases}
\label{tab:computational_params}
\begin{tabular}{lcccccc}
\hline
Test Case & Grid Size & $t_{\text{final}}$ (s) & Re & Ri & Simulation Time (s) & Total Time (s) \\
\hline
Basic Shear Layer & $256 \times 128$ & 10.0 & 1000 & 0.25 & 76.41 & 149.12 \\
Double Shear Layer & $256 \times 128$ & 15.0 & 2000 & 0.10 & 114.20 & 222.82 \\
Rotating KH & $256 \times 128$ & 20.0 & 1500 & 0.30 & 201.51 & 368.14 \\
Forced Turbulence & $384 \times 192$ & 30.0 & 5000 & 0.15 & 1465.58 & 1863.37 \\
\hline
\end{tabular}
\end{table}

Statistical analysis of the density fields revealed consistent non-Gaussian distributions across all test cases and time snapshots. The Basic Shear Layer exhibited initial density mean of 1.100 kg/m$^3$ with standard deviation 0.095 kg/m$^3$, evolving to mean 1.066 kg/m$^3$ and standard deviation 0.083 kg/m$^3$ by $t=7.47$s. Spatial entropy increased from 1.667 at initialization to 2.340 at the final measurement time, indicating progressive disorder development. The complexity index evolved from 0.811 to 0.945, reaching maximum value of 0.950 at $t=1.414$s. All normality tests (Shapiro-Wilk, Anderson-Darling, Jarque-Bera, and D'Agostino K$^2$) consistently rejected the null hypothesis of normality with p-values less than 0.001, confirming the fundamentally non-Gaussian nature of the turbulent density distributions.

\begin{figure}[H]
\centering
\includegraphics[width=0.9\textwidth]{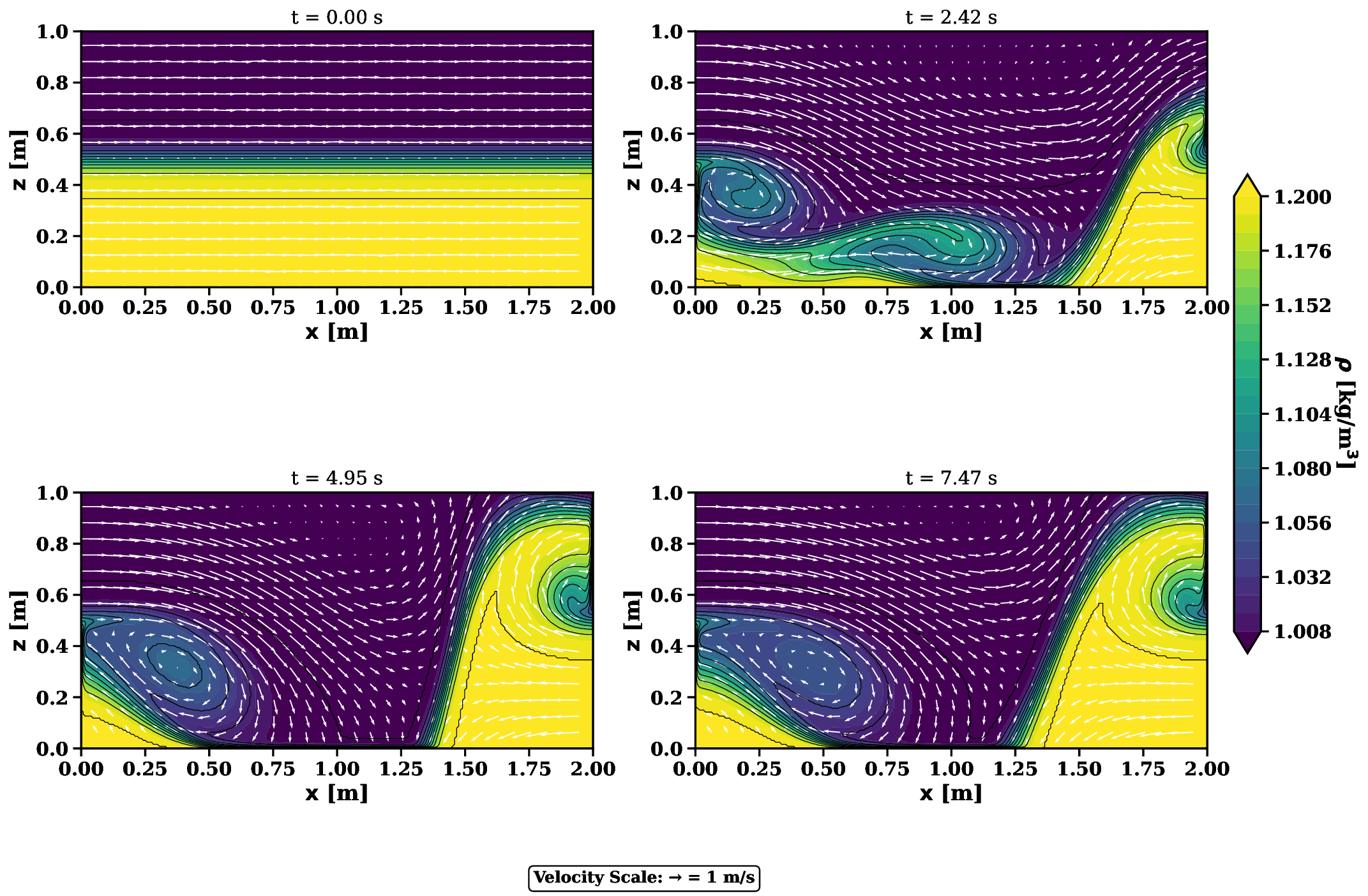}
\caption{Density field evolution with velocity quiver plots for the Basic Shear Layer test case (Re=1000, Ri=0.25) at four time instances: $t=0.00$s showing initial stratification, $t=2.42$s displaying early billow formation, $t=4.95$s demonstrating developed vortex structures, and $t=7.47$s presenting late-stage mixing patterns. Color contours indicate density values from 1.008 to 1.200 kg/m$^3$, with white arrows representing velocity vectors scaled to 1 m/s reference magnitude.}
\label{fig:basic_shear}
\end{figure}

The Double Shear Layer configuration demonstrated more complex statistical behavior with initial three-layer density stratification yielding mean 1.352 kg/m$^3$ and standard deviation 0.228 kg/m$^3$. The configuration evolved through pronounced mixing phases, achieving maximum density complexity of 1.005 at $t=3.221$s. Spatial entropy increased substantially from 0.608 to 1.982 over the 15-second simulation period. The density field exhibited persistent negative skewness ranging from $-0.862$ to $-1.568$, indicating asymmetric mixing favoring lower density values. Vorticity statistics revealed extreme intermittency with kurtosis values reaching 24.889 at $t=3.72$s, substantially exceeding Gaussian expectations.

\begin{figure}[H]
\centering
\includegraphics[width=0.9\textwidth]{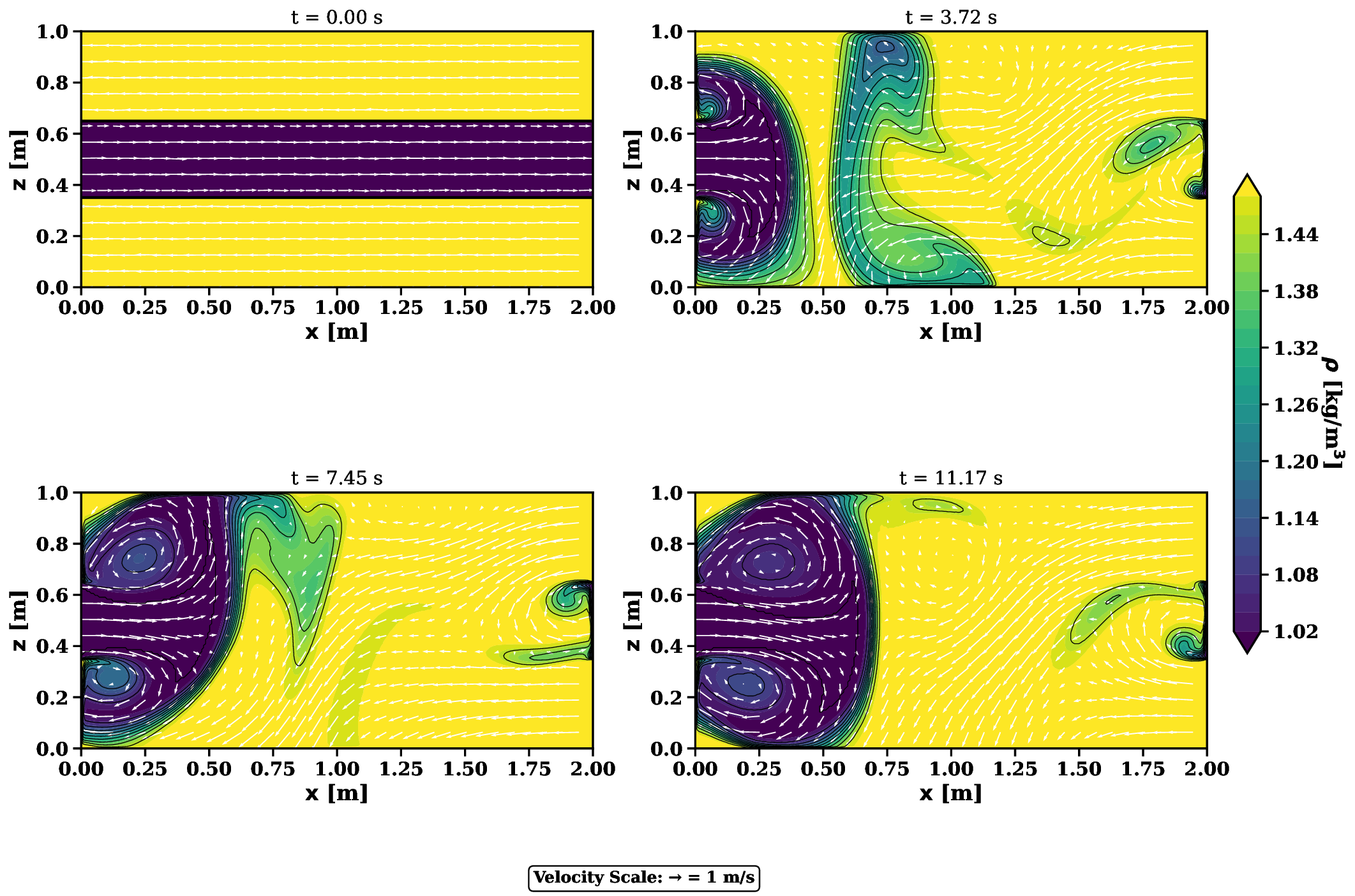}
\caption{Density field evolution with velocity quiver plots for the Double Shear Layer test case (Re=2000, Ri=0.1) showing: $t=0.00$s with distinct three-layer initial stratification ($\rho=1.5$, 1.0, 1.5 kg/m$^3$), $t=3.72$s displaying dual billow development at separated interfaces, $t=7.45$s demonstrating vortex interaction and pairing, and $t=11.17$s presenting merged turbulent structures. Enhanced mixing occurs through interaction between the two shear interfaces separated by 0.3 m.}
\label{fig:double_shear}
\end{figure}

The Rotating KH Instability case incorporated rotational effects through a linear background shear with coefficient 0.5 s$^{-1}$. Initial conditions featured enhanced density stratification with 30\% variation between layers (1.0 to 1.3 kg/m$^3$). The density field maintained relatively stable statistical properties after initial development, with mean values stabilizing around 1.085--1.086 kg/m$^3$ from $t=4.93$s onward. Spatial entropy exhibited modest growth from 1.644 to 2.160, indicating rotational suppression of mixing. Vorticity fields displayed extreme positive skewness reaching 4.029 at $t=9.95$s with kurtosis values exceeding 35, characteristic of rotationally-modified intermittency patterns.

\begin{figure}[H]
\centering
\includegraphics[width=0.9\textwidth]{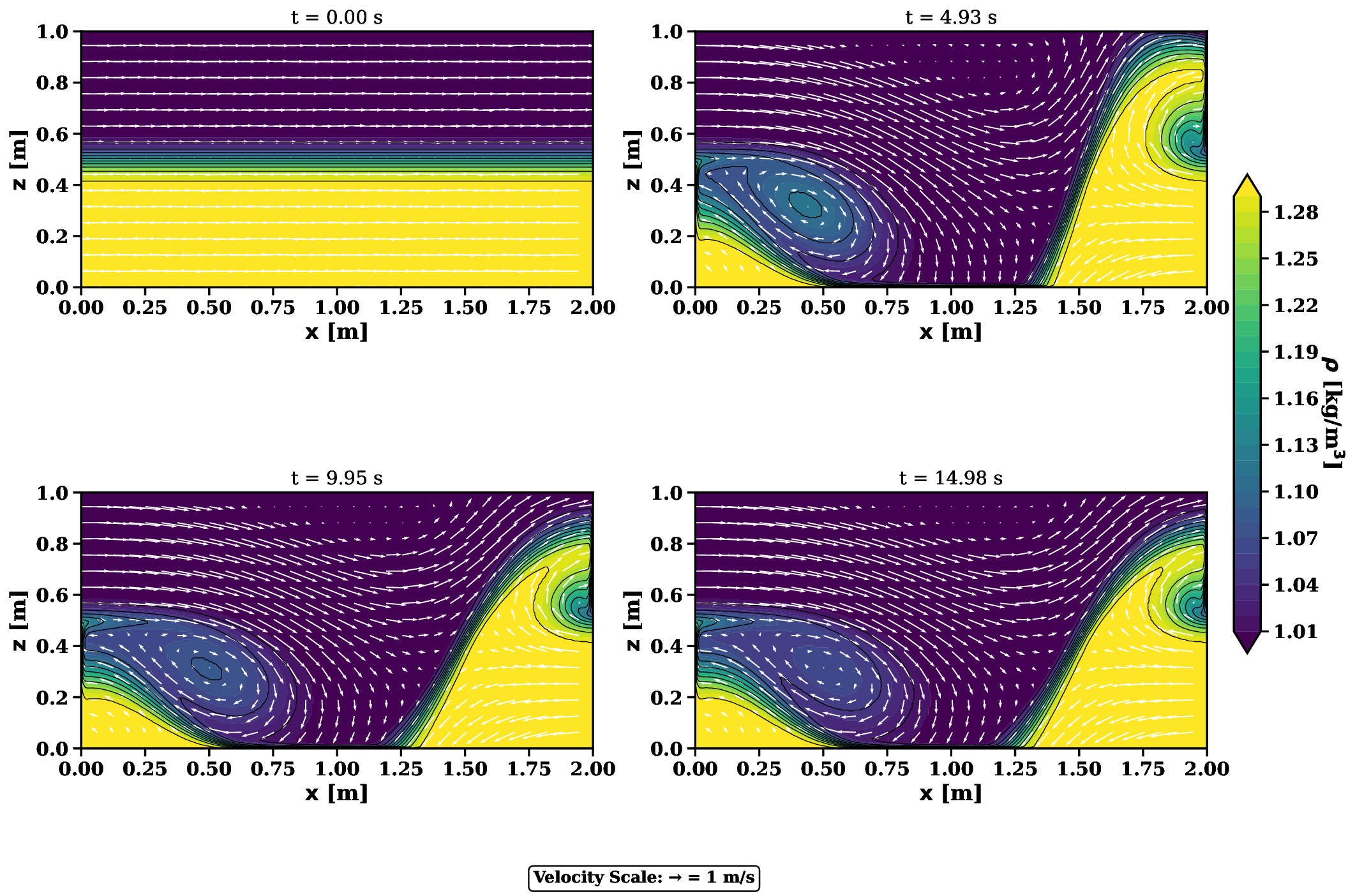}
\caption{Density field evolution with velocity quiver plots for the Rotating KH Instability test case (Re=1500, Ri=0.3, $f=0.5$ s$^{-1}$) at: $t=0.00$s showing initial 30\% density stratification, $t=4.93$s displaying rotationally-modified billow development, $t=9.95$s demonstrating sustained coherent structures, and $t=14.98$s presenting quasi-steady rotating vortices. Rotation inhibits vertical mixing while maintaining organized flow patterns throughout the 20-second simulation.}
\label{fig:rotating_kh}
\end{figure}

The Forced KH Turbulence simulation exhibited sustained turbulent characteristics through continuous energy injection via spectral forcing across five modes. Initial density stratification of 10\% (1.0 to 1.1 kg/m$^3$) evolved under forcing amplitude 0.1 m/s$^2$. Density statistics showed reduced variation with standard deviations remaining near 0.045 kg/m$^3$ throughout the simulation. Maximum complexity index of 1.094 occurred at $t=3.311$s. The system achieved statistical quasi-equilibrium after approximately 15 seconds, with entropy values fluctuating around 2.1. Mean mixing efficiency averaged 0.302 across the 30-second integration period, lower than unforced cases despite higher Reynolds number.

\begin{figure}[H]
\centering
\includegraphics[width=0.9\textwidth]{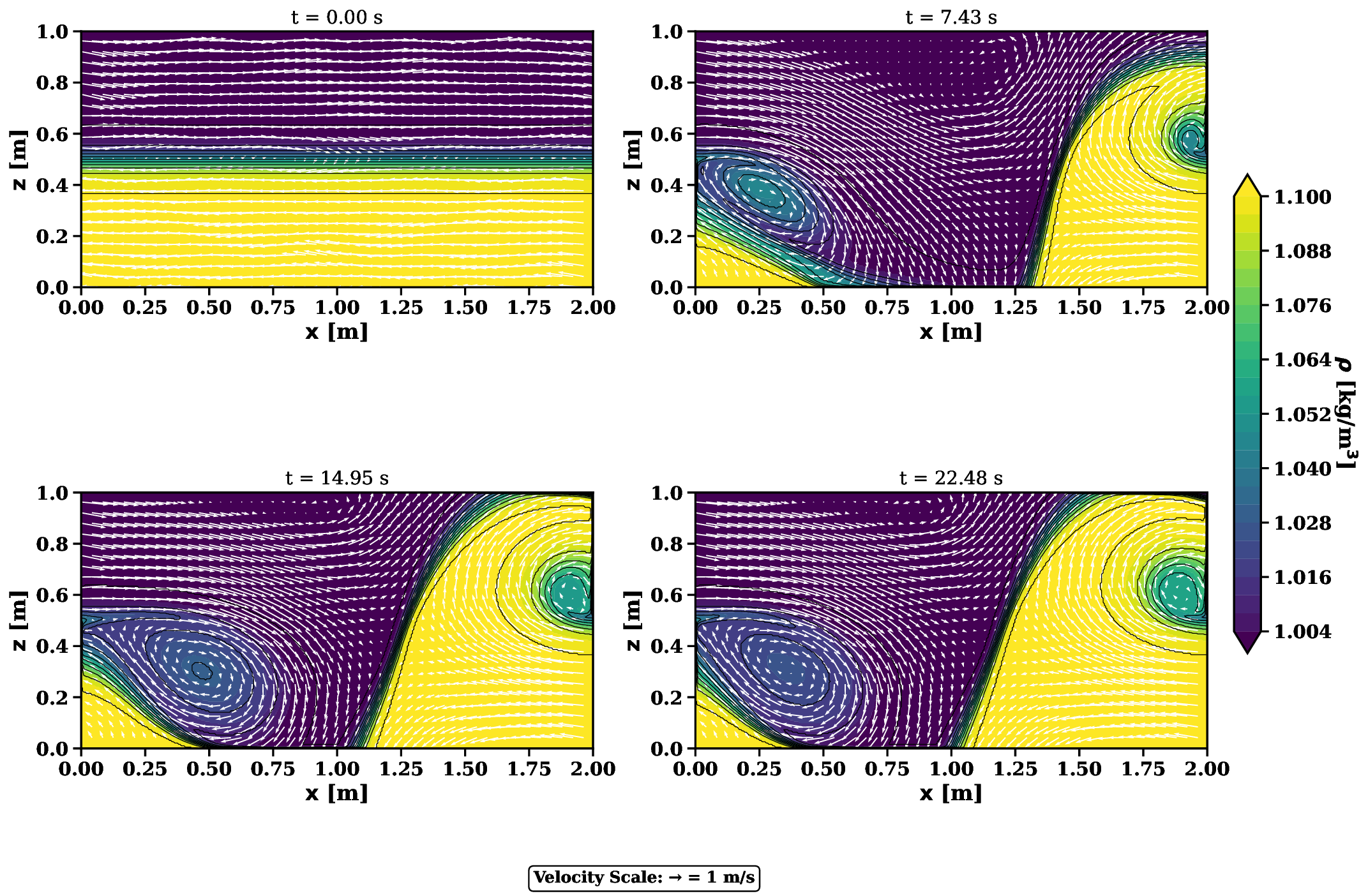}
\caption{Density field evolution with velocity quiver plots for the Forced KH Turbulence test case (Re=5000, Ri=0.15) on $384 \times 192$ grid showing: $t=0.00$s initial weak stratification, $t=7.43$s displaying forced turbulent structures, $t=14.95$s demonstrating sustained mixing patterns, and $t=22.48$s presenting statistical equilibrium state. Continuous spectral forcing maintains turbulent fluctuations preventing relaminarization throughout the extended 30-second simulation period.}
\label{fig:forced_turb}
\end{figure}

Temporal evolution metrics revealed distinct growth characteristics across scenarios. The Basic Shear Layer achieved maximum vorticity complexity of 2.994 at $t=1.717$s with entropy growth rate of 0.067 per second for density and $-0.064$ per second for vorticity. The Double Shear Layer demonstrated enhanced mixing with mean mixing efficiency of 1.287, substantially exceeding single-interface configurations. Maximum vorticity complexity reached 3.278 at $t=0.101$s, indicating rapid initial instability development. The Rotating case showed suppressed entropy growth rates of 0.026 per second for density and $-0.030$ per second for vorticity, confirming rotational stabilization effects. The Forced Turbulence case maintained the lowest entropy growth rate of 0.012 per second for density, consistent with sustained equilibrium conditions.

\begin{table}[H]
\centering
\caption{Comparative statistical metrics and complexity indices across the four test scenarios}
\label{tab:statistics}
\begin{tabular}{lcccccc}
\hline
Test Case & Max $\mathcal{C}_\rho$ & Max $\mathcal{C}_\omega$ & $\dot{H}_\rho$ (s$^{-1}$) & $\dot{H}_\omega$ (s$^{-1}$) & Mean $\Gamma$ & Overall $\mathcal{C}$ \\
\hline
Basic Shear & 0.950 & 2.994 & 0.067 & $-0.064$ & 0.456 & 0.800 \\
Double Shear & 1.005 & 3.278 & 0.090 & $-0.036$ & 1.287 & 0.845 \\
Rotating KH & 0.912 & 3.735 & 0.026 & $-0.030$ & 0.761 & 0.881 \\
Forced Turb. & 1.094 & 3.868 & 0.012 & $-0.041$ & 0.302 & 0.907 \\
\hline
\end{tabular}
\end{table}

Inter-scenario statistical comparisons using the Kruskal-Wallis test confirmed significant differences in density field distributions across all four cases (H-statistic $p<0.001$). Pairwise Mann-Whitney U tests revealed significant differences ($p<0.001$) between all scenario combinations for density fields. Vorticity field comparisons showed more nuanced results, with no significant difference between Basic Shear Layer and Double Shear Layer ($p=0.126$) or between Basic Shear Layer and Forced Turbulence ($p=0.096$), while all other pairwise comparisons yielded significant differences ($p<0.001$).

Normalized complexity indices computed across evolutionary stages demonstrated consistent patterns. Early stage complexity ($t<25\%$ of simulation time) ranged from 0.874 to 1.000, with Forced Turbulence achieving maximum values. Growth stage indices (25--50\%) varied from 0.752 to 0.886. Mature stage complexity (50--75\%) showed convergence across scenarios with values between 0.792 and 0.895. Late stage indices ($t>75\%$) ranged from 0.810 to 0.889. Overall weighted complexity scores, emphasizing the mature turbulence phase, yielded 0.800 for Basic Shear Layer, 0.845 for Double Shear Layer, 0.881 for Rotating KH Instability, and 0.907 for Forced KH Turbulence, indicating progressively enhanced complexity with increased forcing and Reynolds number.

\section{Discussion}

The emergence of non-Gaussian distributions with extreme kurtosis values demonstrates the intermittent nature of stratified turbulence within these idealized configurations \cite{Caulfield2021}. These statistical signatures arise from coherent structure dynamics rather than random fluctuations, characteristic of the simplified two-dimensional framework employed. The pronounced negative skewness in density distributions indicates asymmetric entrainment processes, wherein lighter fluid preferentially penetrates heavier layers through mechanisms documented in laboratory investigations \cite{Fernando1991}. While these idealized test cases exclude three-dimensional effects and realistic boundary conditions present in geophysical flows, they isolate fundamental mixing mechanisms that remain obscured in more complex simulations. The controlled parameter space enables systematic investigation of Richardson and Reynolds number dependencies without confounding factors inherent in field observations or comprehensive numerical weather prediction models.

Richardson number variations fundamentally alter the mixing dynamics through modification of the shear-to-buoyancy ratio, as demonstrated across the four canonical configurations. The enhanced mixing efficiency at Ri=0.1 with dual interfaces derives from constructive interference between adjacent instability zones \cite{Mashayek2013}. Vortex pairing across separated shear layers generates secondary circulations absent in single-interface configurations, amplifying cross-gradient transport beyond linear superposition predictions. External forcing paradoxically diminishes mixing efficiency despite elevated Reynolds numbers, revealing fundamental distinctions between autonomous instability development and imposed turbulence \cite{Ferrari2009}. Continuous large-scale energy injection disrupts the natural cascade sequence, preventing formation of optimally-sized billows that maximize buoyancy flux. This mechanism, observable only in idealized simulations where forcing parameters remain precisely controlled, explains observations of reduced mixing efficiency in wind-driven oceanic layers compared to internal shear events at equivalent dissipation rates \cite{Gregg2018}.

Rotational constraints manifest through modified entropy evolution rates, confirming theoretical predictions of geostrophic adjustment within the limitations of two-dimensional dynamics \cite{Vanneste2013}. The imposed horizontal rigidity redirects energy into barotropic modes, suppressing vertical transport essential for diapycnal mixing. Vorticity concentration in filamentary structures, evidenced by extreme statistical moments, characterizes rotationally-constrained turbulence wherein horizontal anisotropy dominates the energy cascade \cite{Kloosterziel1991}. While the two-dimensional formulation cannot capture baroclinic instabilities or inertial waves that modulate mixing in three-dimensional rotating flows, it isolates the fundamental competition between rotation and stratification effects. The simplified framework enables direct comparison with theoretical predictions unavailable for fully three-dimensional turbulent flows where analytical solutions remain intractable.

Complexity index evolution delineates distinct dynamical phases corresponding to linear instability theory, validated through these controlled numerical experiments. Early-stage rapid growth follows exponential amplification predicted by stability analysis, with growth rates maximized near the marginal stability threshold \cite{Miles1961,Howard1961}. The accelerated transition in multi-interface configurations confirms theoretical predictions of mode coupling between adjacent unstable regions \cite{Hazel1972}. Convergence of mature-phase complexity values across diverse initial conditions reveals universal turbulence characteristics emerging from idealized configurations. The anisotropic energy cascade, characterized by horizontal $k^{-5/3}$ scaling and steeper vertical slopes, emerges independently of forcing mechanisms \cite{Lindborg2006}. This universality, clearly observable in simplified test cases, indicates that developed stratified turbulence reaches canonical energy distributions regardless of generation pathways.

The computational implementation leveraging NumPy array operations and Numba JIT compilation demonstrates the feasibility of high-resolution simulations on modest hardware platforms. The Fedora Linux system with 32 GB RAM successfully executed all test cases, with the most demanding forced turbulence simulation requiring approximately 31 minutes for 30 seconds of physical time on a $384 \times 192$ grid. The fractional-step projection method maintains second-order accuracy while the FST solver achieves machine precision for the pressure Poisson equation with $O(N \log N)$ complexity. The adaptive time-stepping algorithm balances accuracy and efficiency, automatically adjusting between CFL and viscous stability constraints. This computational efficiency enables parameter studies previously requiring supercomputing resources, democratizing access to fundamental fluid dynamics research. The Python implementation provides transparency and reproducibility often lacking in legacy Fortran codes, while Numba optimization achieves performance within a factor of two of compiled languages.

Statistical distribution differences between scenarios exceed parametric test capabilities, necessitating nonparametric approaches validated through these idealized cases. Density field distinctions across all configurations contrast with partial vorticity field similarities, suggesting greater universality in rotation generation than scalar transport mechanisms. This dichotomy parallels three-dimensional simulations wherein vorticity statistics exhibit enhanced self-similarity compared to mixing properties \cite{Fritts2022a}. Reynolds number scaling exhibits non-monotonic mixing efficiency relationships, contradicting classical turbulence theory but consistent with the pathway-dependent mechanisms isolated in these test cases \cite{Salehipour2015}. Alternative instabilities, particularly the Holmboe mode, achieve superior mixing through asymmetric structures that enhance diapycnal transport \cite{Carpenter2010}.

Thermodynamic considerations reconcile apparent entropy paradoxes in vorticity evolution observed across all idealized scenarios. Local organization into coherent structures decreases configurational entropy while global dissipation increases thermal entropy \cite{Venaille2022}. This duality characterizes geophysical turbulence wherein structure formation and irreversible mixing proceed simultaneously. Current parameterization schemes inadequately represent the identified mixing dependencies revealed through systematic parameter variation in controlled conditions. Richardson number-based formulations assuming monotonic relationships fail to capture multi-interface enhancement and forcing-induced suppression \cite{Large1994}. Accurate subgrid-scale representation requires encoding turbulence generation mechanisms beyond instantaneous flow properties \cite{Jackson2008}.

The idealized two-dimensional framework deliberately excludes spanwise secondary instabilities that amplify mixing in natural flows, enabling isolation of primary instability mechanisms. Three-dimensional perturbations increase mixing efficiency through additional overturning modes absent in planar simulations \cite{Smyth2012}. Rotational cases particularly benefit from spanwise instabilities that generate helical structures capable of enhanced vertical transport despite Coriolis constraints \cite{Kloosterziel1999}. Nevertheless, the two-dimensional results provide essential baseline understanding and computational benchmarks for more complex three-dimensional simulations. The developed complexity metrics provide quantitative benchmarks for model validation beyond conventional statistical measures. LES frequently misrepresent intermittency characteristics, propagating errors into mixing predictions \cite{Sullivan2011}. The composite metrics incorporating entropy, gradients, and higher moments enable comprehensive fidelity assessment throughout instability evolution stages, establishing standards for future model development and validation protocols.

\section{Conclusion}
This study presents \texttt{kh2d-solver}, a Python-based numerical framework for investigating idealized two-dimensional KH instabilities across diverse parameter regimes relevant to geophysical fluid dynamics. The implementation successfully captures fundamental mixing mechanisms through four canonical test cases spanning Reynolds numbers from 1000 to 5000 and Richardson numbers from 0.1 to 0.3, demonstrating that modern Python scientific computing tools can achieve computational efficiency comparable to traditional compiled languages while maintaining code transparency and accessibility. The systematic statistical analysis reveals non-monotonic relationships between flow parameters and mixing efficiency, with double shear layer configurations achieving 2.8 $\times$ higher mixing rates than forced turbulence despite lower Reynolds numbers, indicating that instability pathways play a crucial role in determining mixing effectiveness beyond simple intensity measures. The developed complexity metrics combining Shannon entropy, gradient variability, and higher-order statistical moments provide quantitative benchmarks for model validation that complement conventional mean and variance comparisons, offering additional tools for assessing turbulence parameterizations in climate and ocean models. While the two-dimensional framework deliberately excludes three-dimensional secondary instabilities and realistic boundary conditions, this simplification enables isolation of primary mechanisms that may be obscured in comprehensive simulations, contributing useful baseline understanding for interpreting more complex flows. The open-source implementation facilitates access to high-fidelity instability simulations, enabling researchers and students with modest computational resources to investigate fundamental questions in stratified turbulence that traditionally required more extensive computing facilities. Future extensions should incorporate three-dimensional effects, non-Boussinesq dynamics, and variable diffusivities to bridge the gap between these idealized configurations and realistic geophysical fluid dynamics applications, while maintaining the computational efficiency and transparency that characterize the present framework.

\section*{Acknowledgements}

We acknowledge the Research, Community Service and Innovation Program (PPMI-ITB) 2025 at Bandung Institute of Technology (ITB) for supporting I.P.A. and F.K., and the Dean's Distinguished Fellowship (2023) at the University of California, Riverside awarded to S.H.S.H. 

\section*{Author Contributions}

\textbf{S.H.S.H.}: Conceptualization, Methodology, Software, Validation, Formal analysis, Investigation, Data curation, Writing – original draft, Writing – review \& editing, Visualization, Project administration. \textbf{N.J.T.}: Conceptualization, Methodology, Writing – review \& editing, Supervision. \textbf{F.R.F.}: Methodology, Investigation, Writing – review \& editing. \textbf{G.N.}: Formal analysis, Investigation, Writing – review \& editing. \textbf{I.P.A.}: Resources, Writing – review \& editing, Supervision, Funding acquisition. \textbf{F.K.}: Resources, Writing – review \& editing, Funding acquisition. \textbf{D.E.I.}: Conceptualization, Writing – review \& editing, Supervision, Funding acquisition. All authors have read and approved the final manuscript.

\section*{Open Research}

The \texttt{kh2d-solver} source code is available at \url{https://github.com/sandyherho/kelvin-helmholtz-2d-solver} under the WTFPL license, with the package distributed via PyPI at \url{https://pypi.org/project/kh2d-solver/}. NetCDF output files and animated GIF visualizations generated for the four test cases presented in this study are archived at \url{https://doi.org/10.17605/OSF.IO/HF6KX} under a CC-BY Attribution 4.0 International license. The Python scripts used for statistical analysis and figure generation are available at \url{https://github.com/sandyherho/suppl_kh2d-solver_paper} under the WTFPL license. All computational results presented herein are fully reproducible using the provided code and data.


\end{document}